\documentclass[final,3p,times]{elsarticle}
\usepackage{graphicx,setspace}
\usepackage{psfrag}
\usepackage{amsfonts}
\usepackage{graphicx}
\usepackage{amsmath,amssymb}
\usepackage{geometry}
\expandafter\let\csname equation*\endcsname\relax
\expandafter\let\csname endequation*\endcsname\relax
\usepackage{amsthm}
\usepackage{mathrsfs}
\usepackage{epstopdf}
\usepackage{float}
\usepackage{lineno,hyperref}
\usepackage{color}
\usepackage{subfigure}
\usepackage{amsmath}
\usepackage{dsfont} 
\usepackage{amssymb} 
\usepackage{graphicx,color}
\usepackage{extarrows}

\usepackage{graphicx}

\usepackage{epstopdf}

\journal{Arxiv}

\begin{document}
\newtheorem{definition}{Definition}[section]
\newtheorem{lemma}{Lemma}[section]
\newtheorem{remark}{Remark}[section]
\newtheorem{theorem}{Theorem}[section]
\newtheorem{proposition}{Proposition}
\newtheorem{assumption}{Assumption}
\newtheorem{example}{Example}
\newtheorem{corollary}{Corollary}[section]
\def\ep{\varepsilon}
\def\Rn{\mathbb{R}^{n}}
\def\Rm{\mathbb{R}^{m}}
\def\E{\mathbb{E}}
\def\hte{\hat\theta}
\renewcommand{\theequation}{\thesection.\arabic{equation}}
\begin{frontmatter}

\title{Critical Ising Model on the Periodic Cubic Lattice
	\((\mathbb{Z}/L\mathbb{Z})^{3}\)}
\author{Shenglan Yuan\corref{cor1}}
\ead{shenglanyuan@gbu.edu.cn}
\cortext[cor1]{Corresponding author}
\address[addr1]{Department of Mathematics, School of Sciences, Great Bay University, Dongguan 523000, China}

\begin{abstract}
We investigate the critical behaviour of the nearest-neighbour Ising model on the
three-dimensional periodic cubic lattice
\((\mathbb{Z}/L\mathbb{Z})^{3}\), and we present a companion set of
Monte Carlo simulations.  After recalling the rigorous results on the
continuity of the spontaneous magnetization in \(d=3\) and the triviality
of the scaling limit in \(d\ge 4\), we compile modern high-precision estimates of
the 3D Ising critical exponents and verify the scaling relations that connect them.
We then describe numerical measurements of the Binder cumulant, the
magnetic susceptibility, the specific heat, the two-point correlation
function, and the finite-size scaling of the magnetization.  We discuss the
systematic uncertainties that affect the extraction of the exponent ratios
\(\gamma/\nu\) and \(\alpha/\nu\), and we emphasize that an exponential fit
of the correlation function is not appropriate at the critical point, where
the decay is algebraic.
\end{abstract}


\end{frontmatter}

\section{Introduction}
The Ising model is the canonical lattice model of a ferromagnetic phase
transition.  Spins \(\sigma_{x}\in\{+1,-1\}\) sit on the sites of a
\(d\)-dimensional lattice and interact through a nearest-neighbour
ferromagnetic coupling; at zero external field, alignment of spins minimises the energy, while the \(\mathbb{Z}_{2}\) spin-flip symmetry is restored and entropy is maximised in the disordered phase.  The competition between these two tendencies
produces a continuous order--disorder transition in dimensions \(d\ge 2\)
and a disordered phase at all temperatures in \(d=1\).  Despite the
simplicity of its definition, the model captures the universal behaviour
of a large class of systems near criticality and has become a testing
ground for both rigorous statistical mechanics and high-precision
numerical methods.

In two dimensions the model is exactly solvable.  Onsager's solution of
the square-lattice model established the existence of a critical point and
provided the first analytic values of its critical exponents, and the
subsequent work of Chelkak and Smirnov \cite{Universality} proved that the
critical 2D Ising model is conformally invariant, so that its scaling
limit is described by a conformal field theory with central charge
\(c=1/2\).  In dimensions \(d\ge 4\) the situation is also well
understood, though for the opposite reason: Aizenman \cite{Geometric}
showed that the scaling limit is Gaussian for \(d\ge 5\), and Aizenman and
Duminil-Copin \cite{Marginal} extended the triviality result to the
marginal case \(d=4\).  In both regimes, the critical scaling limit is rigorously understood.

The three-dimensional case occupies the intermediate regime in which the
model is neither integrable nor mean-field.  Aizenman, Duminil-Copin, and
Sidoravicius \cite{Random} proved that the spontaneous magnetization is
continuous at the critical point, so that the transition is continuous
rather than first order, and this result is now part of a broader
understanding of the 3D Ising universality class.  Nevertheless, the
model admits no exact closed-form solution for its critical temperature or
its critical exponents, and a rigorous construction of its scaling limit
remains open.  The conjectural description of that limit by a
three-dimensional Ising conformal field theory, together with the
conformal-bootstrap determination of the exponents \cite{SimmonsDuff},
has made the 3D Ising model one of the most precisely studied
non-integrable models in statistical physics.  See \cite{Multipleequilibria}
for a broader discussion of the role of the model in the theory of phase
transitions.

The precision of the conformal-bootstrap results has raised the standard
for numerical work.  The critical temperature of the simple cubic lattice
is known to high precision,
$$
K_{c}=\frac{J}{k_{B}T_{c}}=0.221654626(5),
$$
and the exponents are known to comparable accuracy,
$$
\alpha\simeq 0.11008,\quad
\beta_{\mathrm{mag}}\simeq 0.326419,\quad
\gamma\simeq 1.237075,\quad
\nu\simeq 0.629971,\quad
\eta\simeq 0.036298,\quad
\delta\simeq 4.78984.
$$
Reproducing these values via Monte Carlo simulation is nontrivial: near \(T_{c}\), the correlation length diverges, and the Metropolis algorithm suffers from critical slowing down. The
extraction of exponents from finite-size scaling is complicated by
corrections to scaling and by the analytic background of the specific
heat.  Cluster algorithms, histogram reweighting, parallel tempering, and
data collapse have become the standard tools for this problem.

The present paper has two aims.  The first is expository: we collect the
rigorous results on the continuity of the transition and on the
triviality of the scaling limit in \(d\ge 4\), the numerical values of the
critical exponents, and the scaling relations that relate them, in a form
that is convenient for readers entering the field.  The second is
numerical: we present a companion set of Monte Carlo simulations of the
nearest-neighbour Ising model on the periodic cubic lattice
\((\mathbb{Z}/L\mathbb{Z})^{3}\), using the Metropolis, Wolff, and
Swendsen--Wang algorithms, and we discuss the measurement of the Binder
cumulant, the magnetic susceptibility, the specific heat, the two-point
correlation function, and the finite-size scaling of the magnetization.
We pay particular attention to two methodological points that are
sometimes overlooked.  The first is that the Hamiltonian on a periodic
lattice must include the boundary-crossing bonds; omitting them
corresponds to open, not periodic, boundary conditions.  The second is
that an exponential fit of the correlation function is valid only in the
disordered phase, while at criticality the decay is algebraic and the
correlation length must be extracted by the second-moment method.

The remainder of the paper is organised as follows. Section \ref{TDIM} investigates the three-dimensional Ising model.
Subsection~\ref{sec:model} defines the Ising model on the periodic cubic
lattice and fixes the notation for the Gibbs measure and the observables.
Subsection~\ref{sec:tc} recalls the critical temperature and the mean-field
comparison.  Subsection~\ref{sec:critical} states the critical exponents and
the scaling relations they satisfy. Subsection~\ref{sec:phases}
summarises the phase structure.  Subsection~\ref{sec:universality} discusses
universality and the conjectural scaling limit.
Section~\ref{sec:methods} describes the numerical methods, and
Section~\ref{sec:results} presents the results, including the Binder
cumulant, the susceptibility and specific-heat finite-size scaling, the
correlation function, and the data collapse of the magnetization.
Section~\ref{sec:discussion} outlines the limitations of the present
study and the improvements that would be needed for a precision
determination of the exponents.

\section{Three-dimensional Ising model}\label{TDIM}
\subsection{Ising model on the periodic cubic lattice}\label{sec:model}
Let \(L\ge 3\) be an integer and let
$$
\Lambda \;=\; (\mathbb{Z}/L\mathbb{Z})^{3}
$$
be the periodic cubic lattice of linear size \(L\), so that
\(|\Lambda|=L^{3}=N\).  Ising spins are
$$
\sigma_{x}\in\{+1,-1\},\qquad x\in\Lambda .
$$
The Hamiltonian in a magnetic field \(h\) is
\begin{equation}
	\label{eq:H}
	H_{\Lambda,h}(\sigma)
	\;=\;
	-J\sum_{\langle xy\rangle_{\mathrm{per}}}\sigma_{x}\sigma_{y}
	\;-\;
	h\sum_{x\in\Lambda}\sigma_{x},
\end{equation}
where \(J>0\) is the ferromagnetic coupling and
\(\langle xy\rangle_{\mathrm{per}}\) denotes the set of all
nearest-neighbour pairs of the torus, including the wrapped bonds that
connect opposite faces of the cube.  Equivalently, for \(L\ge 3\),
$$
H_{\Lambda,h}(\sigma)
\;=\;
-J\sum_{x\in\Lambda}\sum_{\mu=1}^{3}
\sigma_{x}\,\sigma_{x+\hat\mu}
\;-\;
h\sum_{x\in\Lambda}\sigma_{x},
$$
where the shifted site \(x+\hat\mu\) is understood modulo \(L\):
$$
x+\hat\mu \pmod L,
$$
and the sum over \(\mu=1,2,3\) runs over the three positive coordinate
directions.  With this convention each undirected bond of the torus is
counted exactly once.  The coordination number is \(z=6\) for \(L\ge 3\);
the case \(L=2\) requires a separate convention because positive and
negative directions coincide.

The Gibbs measure at inverse temperature \(\beta=1/(k_{B}T)\) is
$$
\mu_{\Lambda,\beta,h}(\sigma)
=
\frac{1}{Z_{\Lambda}(\beta,h)}\,
e^{-\beta H_{\Lambda,h}(\sigma)},
$$
with the partition function
$$
Z_{\Lambda}(\beta,h)
=
\sum_{\sigma\in\{\pm1\}^{\Lambda}}
e^{-\beta H_{\Lambda,h}(\sigma)} .
$$
The magnetization density is
$$
m(\beta,h)
=
\lim_{\Lambda\uparrow\mathbb{Z}^{3}}
\frac{1}{|\Lambda|}
\Big\langle \sum_{x\in\Lambda}\sigma_{x}\Big\rangle_{\Lambda,\beta,h},
$$
where the limit is taken in the thermodynamic sense. For \(h\neq0\), this limit is unique and independent of boundary conditions. At \(h=0\), the infinite-volume Gibbs measure may be a mixture of pure states; the spontaneous magnetization \(m_*(\beta)\) is defined by symmetry-breaking boundary conditions or by taking \(h\downarrow0\) after the thermodynamic limit:
$$
m_{*}(\beta)=\lim_{h\downarrow 0}m(\beta,h).
$$
This limit yields the positive pure-state magnetization; for \(T<T_c\), the zero-field infinite-volume Gibbs state corresponds to a symmetric convex combination of the two pure states with magnetizations \(\pm m_*\).

\subsection{Critical point}\label{sec:tc}
For \(h=0\) the model has a ferromagnetic phase transition at a critical
inverse temperature \(\beta_{c}\).  For the simple cubic lattice
\(\mathbb{Z}^{3}\),
$$
K_{c}\equiv\frac{J}{k_{B}T_{c}}=0.221654626(5),
$$
so that
$$
T_{c}\simeq 4.51154\,\frac{J}{k_{B}},
\qquad
T_{c}\simeq 4.5115\,J \quad (k_{B}=1).
$$
The transition is second order.  Unlike the two-dimensional Ising model,
the 3D Ising model has no exact closed-form solution for \(T_{c}\) or
for its critical exponents.

Mean-field theory gives
$$
T_{c}^{\mathrm{MF}}=\frac{zJ}{k_{B}}=\frac{6J}{k_{B}},
$$
which overestimates the true \(T_{c}\) by about \(33\%\).

\subsection{Critical behaviour}\label{sec:critical}
Define the reduced temperature
$$
t=\frac{T-T_{c}}{T_{c}} .
$$
At zero field, with \(+\) boundary conditions understood, the spontaneous
magnetization satisfies
$$
m_{*}(\beta)=0 \quad (T>T_{c}),
\qquad
m_{*}(\beta)=+A\,(-t)^{\beta_{\mathrm{mag}}} \quad (T<T_{c}),
$$
with the magnetization exponent \cite{SimmonsDuff}
$$
\beta_{\mathrm{mag}}\simeq 0.326419(3).
$$
If one instead wants to emphasize the two pure phases, one may write
\(m_{*}(\beta)=\pm A(-t)^{\beta_{\mathrm{mag}}}\).  At the critical
temperature,
$$
m(\beta_{c},h)\sim|h|^{1/\delta}\operatorname{sgn}(h),
\qquad
\delta\simeq 4.78984(1).
$$
The other standard critical exponents in \(d=3\) are
$$
\alpha\simeq 0.11008(1),\quad
\gamma\simeq 1.237075(10),\quad
\nu\simeq 0.629971(4),\quad
\eta\simeq 0.036298(2),
$$
governing
$$
C\sim|t|^{-\alpha},\qquad
\chi\sim|t|^{-\gamma},\qquad
\xi\sim|t|^{-\nu},
$$
where \(C\), \(\chi\), \(\xi\) denote the specific heat, the magnetic
susceptibility, and the correlation length.

At \(T=T_{c}\) the correlation length diverges, and the connected two-point
function decays algebraically:
$$
G(x,y)
=
\langle\sigma_{x}\sigma_{y}\rangle
-\langle\sigma_{x}\rangle\langle\sigma_{y}\rangle
\sim
\frac{1}{|x-y|^{d-2+\eta}}
=
\frac{1}{|x-y|^{1+\eta}},
$$
with \(d=3\) and \(1+\eta\simeq 1.0363\).  The system is scale invariant
at criticality; its scaling limit is conjecturally conformally invariant.

The exponents satisfy the scaling relations
$$
\alpha+2\beta_{\mathrm{mag}}+\gamma=2,\qquad
\gamma=\nu(2-\eta),\qquad
2-\alpha=d\nu=3\nu,
$$
and
$$
\delta=\frac{d+2-\eta}{d-2+\eta}=\frac{5-\eta}{1+\eta}.
$$

\subsection{Phase structure}\label{sec:phases}
\begin{itemize}
	\item For \(T<T_{c}\), \(h=0\): ferromagnetically ordered phase.  The
	\(\mathbb{Z}_{2}\) symmetry \(\sigma_{x}\to-\sigma_{x}\) is spontaneously
	broken, and \(m_{*}(\beta)=+m_{0}\neq 0\) with \(+\) boundary conditions,
	which select the positive branch.
	\item For \(T>T_{c}\), \(h=0\): paramagnetic disordered phase with
	\(m_{*}(\beta)=0\) and exponential decay of correlations.
	\item At \(T=T_{c}\), \(h=0\): second-order phase transition.
	In the infinite-volume limit, the free energy becomes non-analytic at \(T_c\); in finite volume all observables remain analytic, and the singularity appears only as a sharp peak.
	The susceptibility diverges, and the 3D Ising
	critical exponents apply.
	\item For \(h\neq 0\): no sharp phase transition; the magnetization is
	nonzero for all temperatures and the transition is smoothed into a
	crossover.
\end{itemize}

\subsection{Universality and scaling limit}\label{sec:universality}
The critical Ising model on \(\mathbb{Z}^{3}\) belongs to the 3D Ising
universality class.  Its long-distance scaling limit is conjecturally
described by the 3D Ising conformal field theory.  The critical
exponents are universal: they do not depend on microscopic details such as
the lattice type, while the critical temperature does.  For example, the
simple cubic, body-centred cubic, and face-centred cubic lattices have
different \(T_{c}\) but the same critical exponents.

\section{Numerical methods}\label{sec:methods}
All simulations reported below use the Hamiltonian \eqref{eq:H} on the
periodic lattice \(\Lambda=(\mathbb{Z}/L\mathbb{Z})^{3}\), with \(J=1\) and
\(k_{B}=1\).  Unless otherwise stated, we use the Wolff single-cluster
algorithm, which substantially reduces critical slowing down near
\(T_{c}\approx 4.5115\).  Observables are sampled only after discarding
\(n_{\mathrm{therm}}\) thermalization sweeps to equilibrate the system.  Statistical errors are
estimated by binning; integrated autocorrelation times are monitored to
confirm that bins are large enough.

The observables we measure are:
\begin{align*}
	m &= \frac{1}{N}\sum_{x\in\Lambda}\sigma_{x},
	&
	e &= -\frac{J}{N}\sum_{\langle xy\rangle_{\mathrm{per}}}\sigma_{x}\sigma_{y},
	\\
	C_{V} &= N\beta^{2}\big(\langle e^{2}\rangle-\langle e\rangle^{2}\big),
	&
	\chi &= N\beta\big(\langle m^{2}\rangle-\langle m\rangle^{2}\big),
	\\
	U_{4} &= 1-\frac{\langle m^{4}\rangle}{3\,\langle m^{2}\rangle^{2}}.
\end{align*}
Here \(e\) and \(m\) are the energy and magnetization per site, and
\(N=L^{3}\).  The lattice sum in the definition of \(e\) runs over the
periodic torus, so that each bond is counted once.  Equivalently, for
\(L\ge 3\),
$$
e
=
-\frac{J}{N}
\sum_{x\in\Lambda}\sum_{\mu=1}^{3}
\sigma_{x}\sigma_{x+\hat\mu},
$$
where the sum over \(\mu=1,2,3\) runs over the three positive coordinate
directions, so that each undirected bond is counted exactly once.

At zero external field \(\langle m\rangle=0\), so the standard finite-size
susceptibility reduces to \(\chi=N\beta\langle m^{2}\rangle\).  Some
studies use the variant
\(\chi=N\beta(\langle m^{2}\rangle-\langle|m|\rangle^{2})\), which
subtracts the square of the absolute magnetization; this variant gives a
smaller amplitude but the same leading scaling exponent.  In this work we
use the standard definition unless otherwise stated.

\section{Numerical results}\label{sec:results}
This section presents the numerical measurements obtained with the
algorithms described in Section~\ref{sec:methods}.  We report the
temperature dependence of the magnetization and the energy
(Subsection~\ref{sec:results-lowres}), the Binder cumulant and the
resulting estimate of \(T_{c}\) (Subsection~\ref{sec:results-binder}), the
finite-size scaling of the susceptibility (Subsection~\ref{sec:results-chi}),
the correlation length extracted from the two-point function
(Subsection~\ref{sec:results-xi}), the finite-size scaling of the specific
heat (Subsection~\ref{sec:results-C}), the data collapse of the
magnetization (Subsection~\ref{sec:results-collapse}), and a set of
supplementary single-point measurements.  Unless otherwise stated, all
measurements use \(J=1\), \(k_{B}=1\), and periodic boundary conditions on
\((\mathbb{Z}/L\mathbb{Z})^{3}\).

\subsection{Low-resolution scans}\label{sec:results-lowres}
\begin{figure}[htbp]
	\centering
	\includegraphics[width=\linewidth]{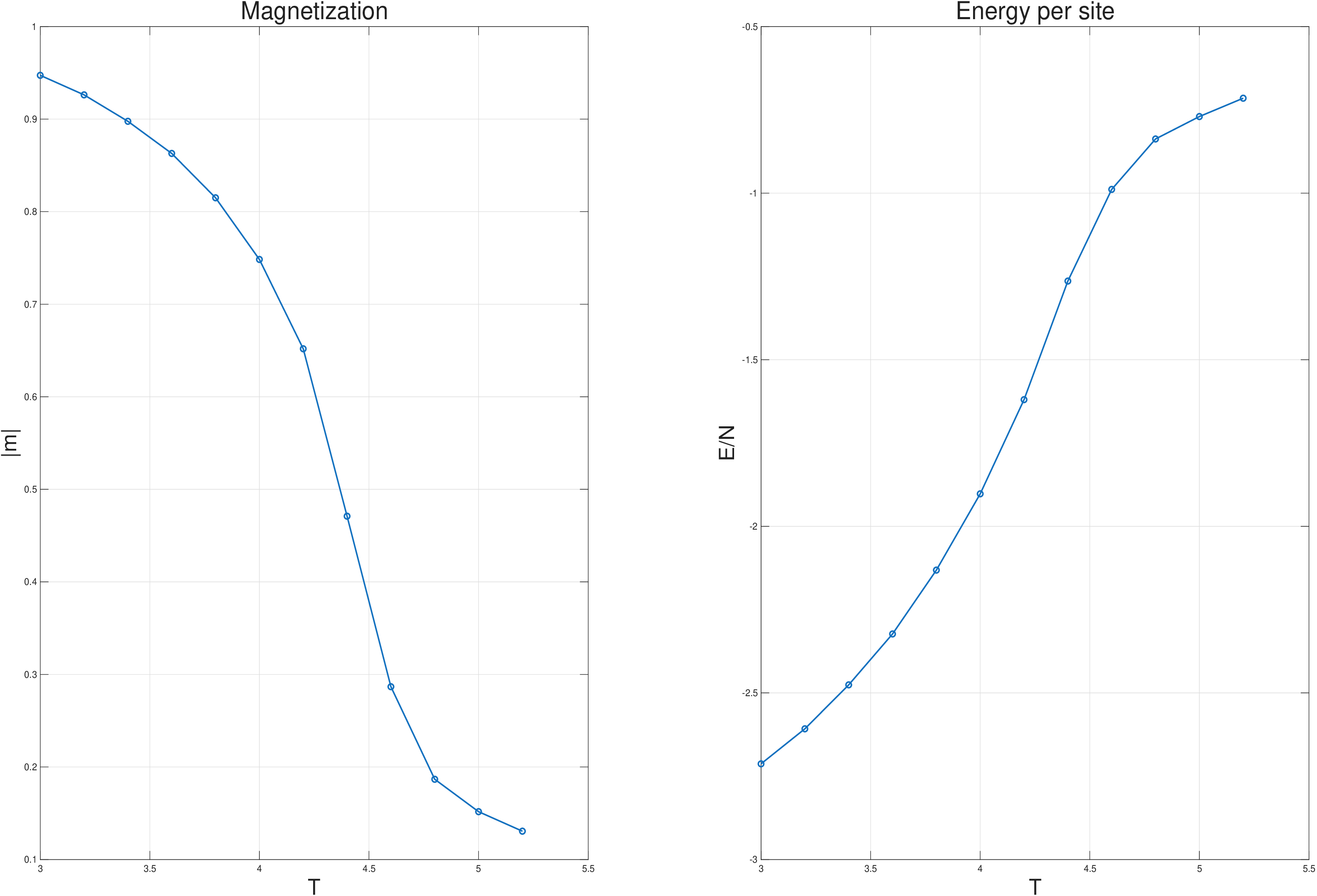}
	\caption{Magnetization \(|m|\) and energy per site \(e\) versus temperature
		for \(L=8\) (\(8^{3}=512\) spins), using the Metropolis algorithm with
		\(T\in[3,5.2]\), step size \(\Delta T=0.2\), \(n_{\mathrm{sweeps}}=2000\), and
		\(n_{\mathrm{therm}}=1000\).  Error bars are smaller than the markers.}
	\label{Code1}
\end{figure}

\begin{figure}[htbp]
	\centering
	\includegraphics[width=\linewidth]{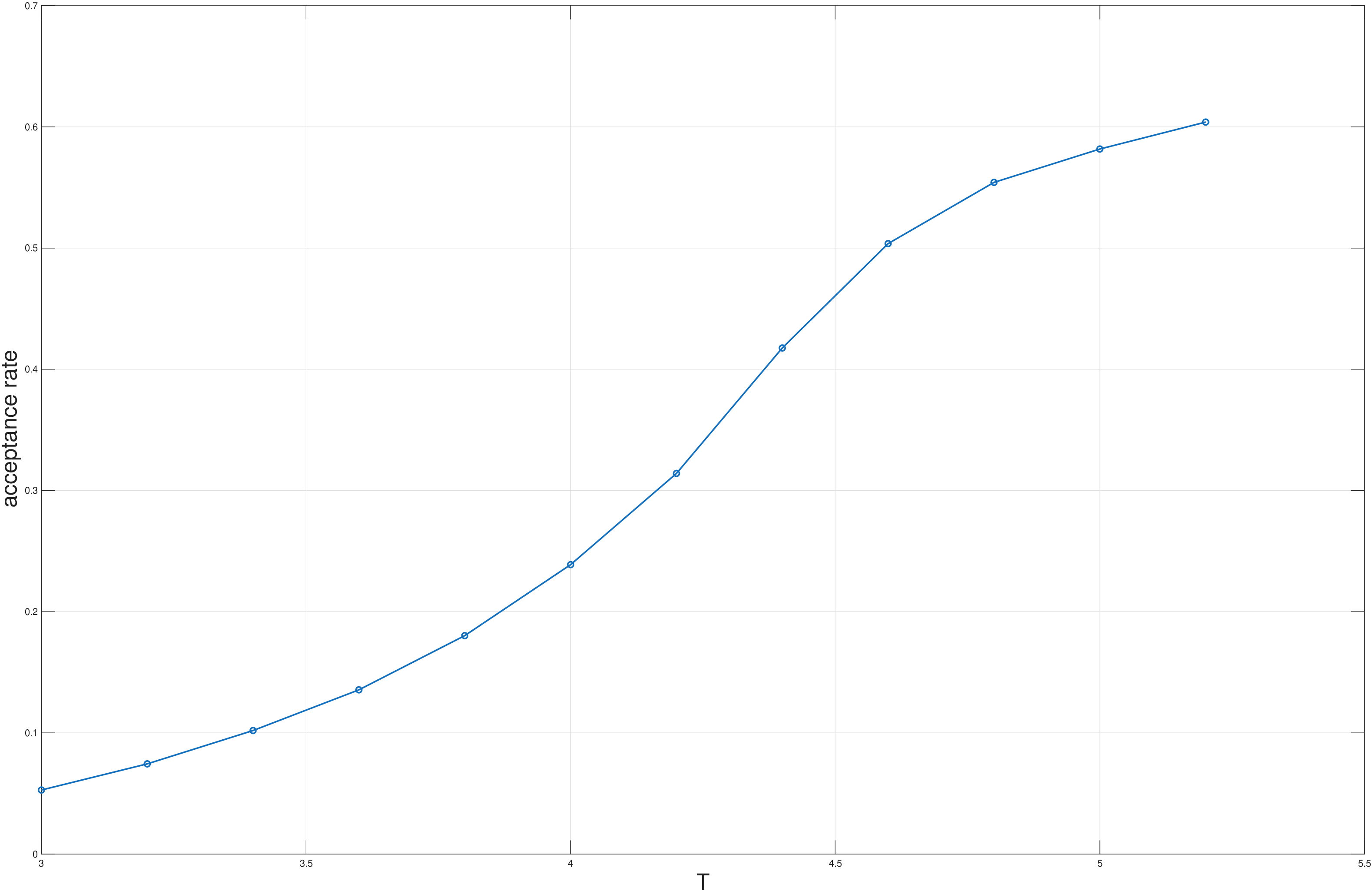}
	\caption{Acceptance rate of Metropolis moves during the measurements shown in Figure~\ref{Code1}. Simulation parameters identical to Figure~\ref{Code1}.}
	\label{Code12}
\end{figure}

Figures~\ref{Code1} and \ref{Code12} show the temperature
dependence of the magnetization and the energy per site for \(L=8\),
obtained with the Metropolis algorithm.  The curves are consistent with a
continuous transition in the vicinity of \(T_{c}\approx 4.5115\).

\begin{figure}[htbp]
	\centering
	\includegraphics[width=\linewidth]{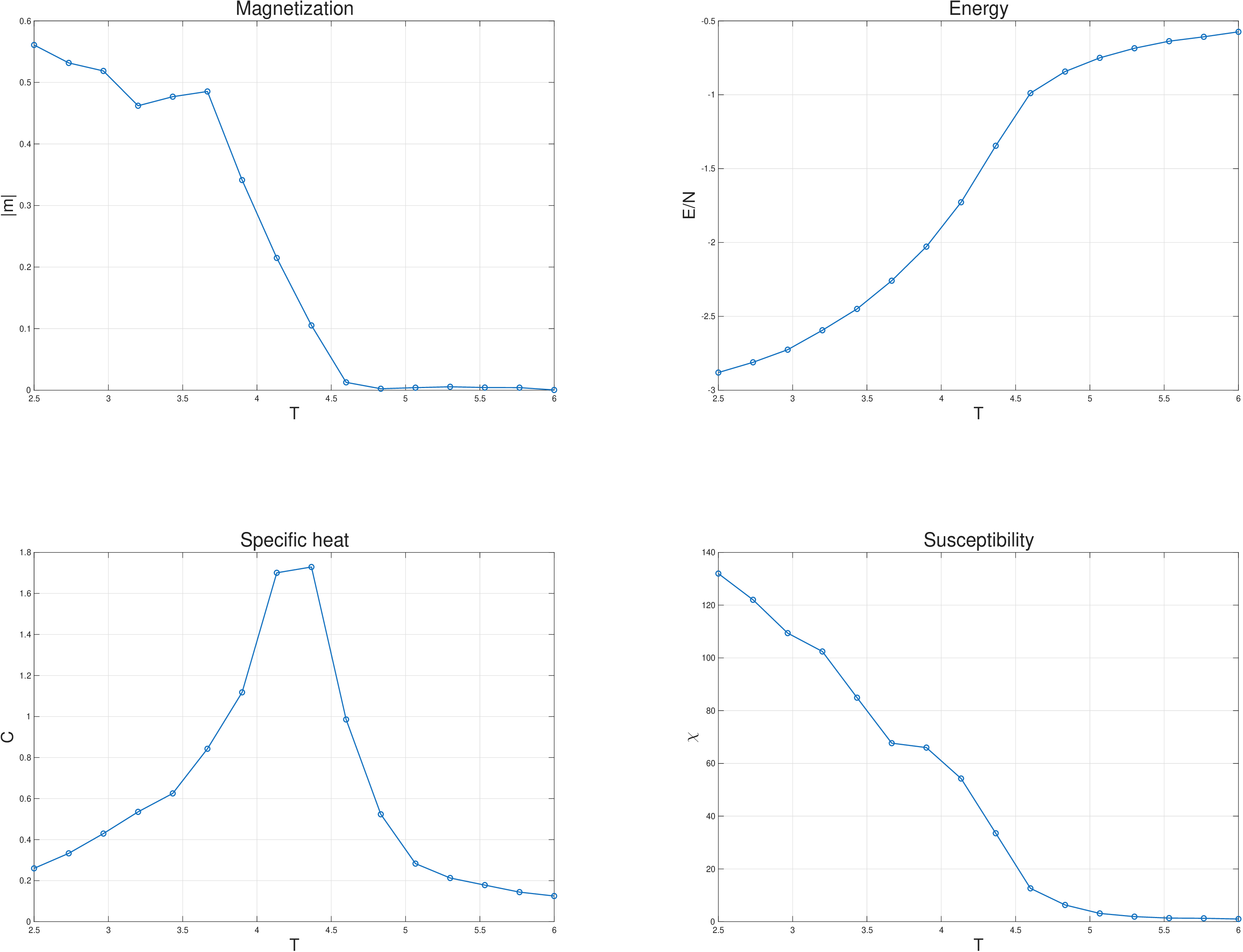}
	\caption{Magnetization, energy, specific heat, and susceptibility measured after thermalization for \(L=8\), obtained by parallel tempering with 16 replicas over \(T\in[2.5,6]\).}
	\label{Code10}
\end{figure}

Figure~\ref{Code10} measures magnetization, energy, specific heat, and susceptibility after thermalization  for \(L=8\), obtained with a parallel-tempering simulation that
uses \(R=16\) replicas with temperatures in the interval \(T\in[2.5,6]\)
and replica exchange between adjacent temperatures.  For each replica a
Metropolis sweep is performed, after which replica exchange is attempted.
Magnetization, energy, specific heat, and susceptibility are measured
after thermalization.

\begin{figure}[htbp]
	\centering
	\includegraphics[width=\linewidth]{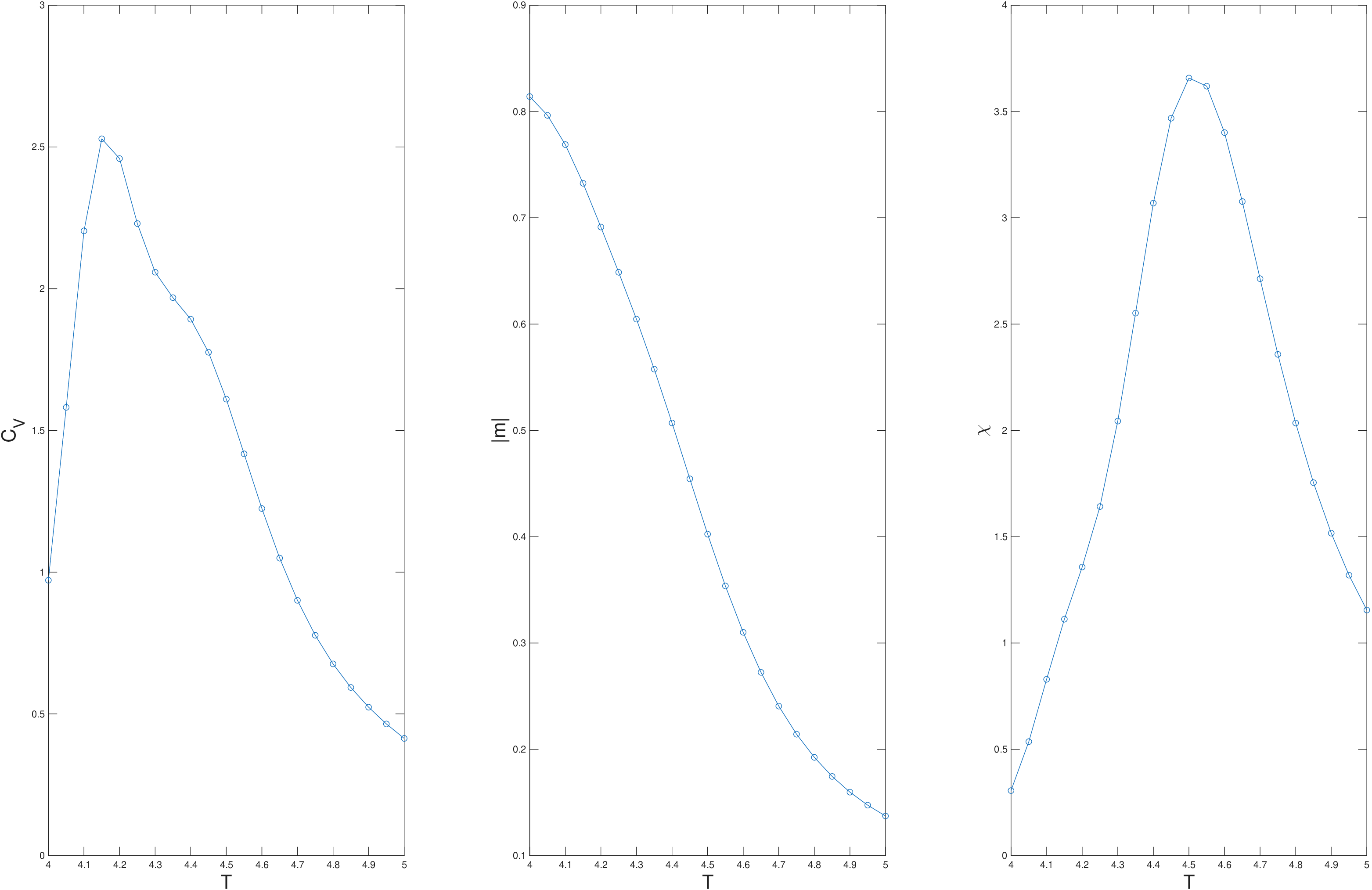}
	\caption{Per-site energy \(e\) and magnetization \(m\) for
		\(L=8\) (\(8^{3}=512\) spins), together with the derived specific heat
		\(C_{V}=N\beta^{2}(\langle e^{2}\rangle-\langle e\rangle^{2})\) and
		susceptibility
		\(\chi=N\beta(\langle m^{2}\rangle-\langle m\rangle^{2})\), using
		\(T\in[4,5]\), step size \(\Delta T=0.05\), \(n_{\mathrm{sweeps}}=5000\), and
		\(n_{\mathrm{therm}}=500\).}
	\label{Code8}
\end{figure}

Figure~\ref{Code8} shows results from single-histogram reweighting at
\(L=8\) for a range of temperatures.  The per-site energy \(e\) and
magnetization \(m\) are shown together with the derived specific heat
\(C_{V}\) and susceptibility \(\chi\).

\begin{figure}[htbp]
	\centering
	\includegraphics[width=\linewidth]{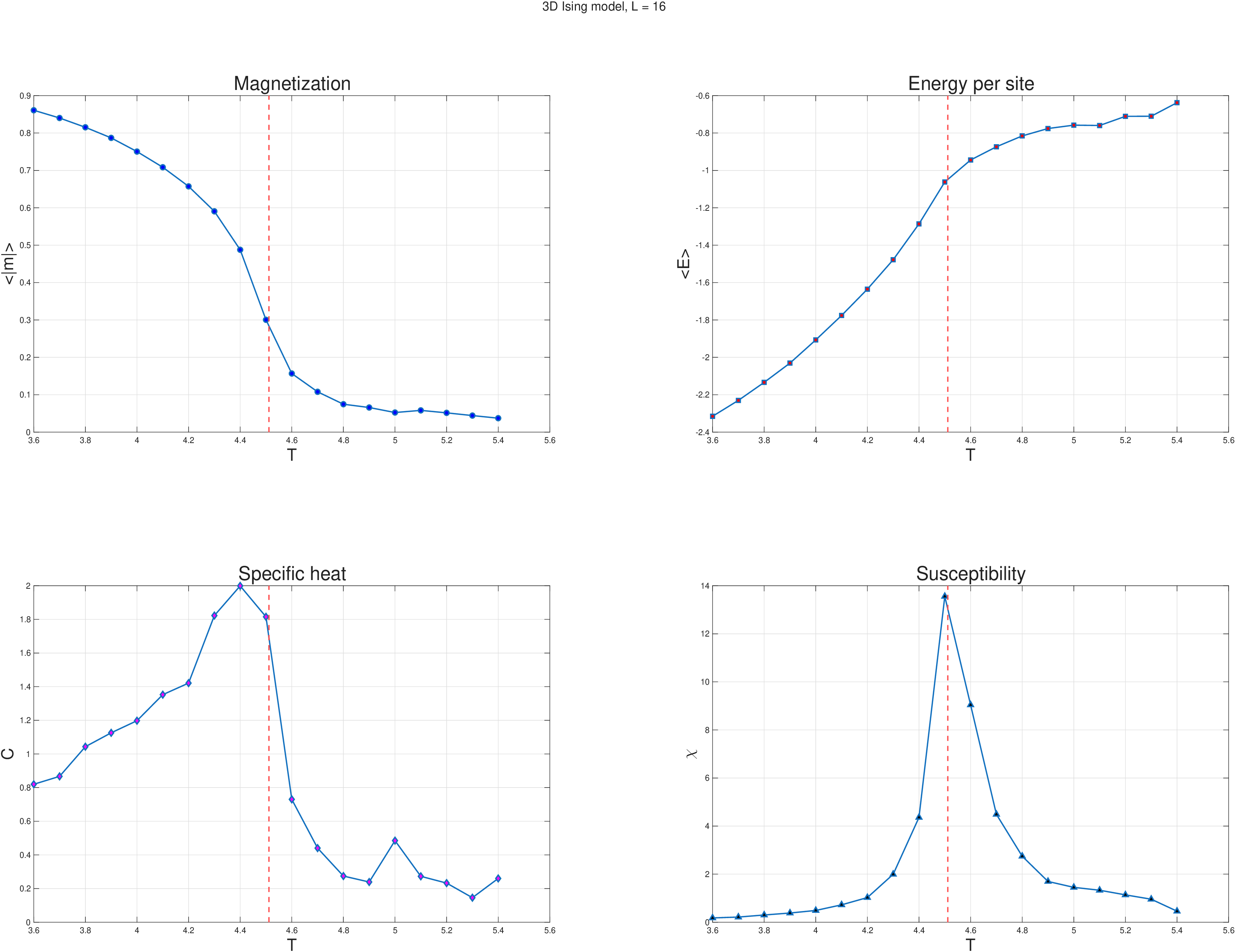}
	\caption{Magnetization \(|m|\), energy per site \(e\), specific heat and susceptibility versus temperature
		for \(L=16\) (\(16^{3}=4096\) spins), using the Wolff single-cluster
		algorithm with \(T\in[3.6,5.4]\), step size \(\Delta T=0.1\), \(4000\) Wolff updates per temperature,
		and \(500\) thermalization updates.  The transition sharpens relative to
		\(L=8\), as expected from finite-size scaling.}
	\label{Code2}
\end{figure}

Figure~\ref{Code2} presents results from a temperature sweep using the Wolff
single-cluster algorithm.  The per-site magnetization and energy are
returned, from which the specific heat and susceptibility are computed.

\begin{figure}[htbp]
	\centering
	\includegraphics[width=\textwidth]{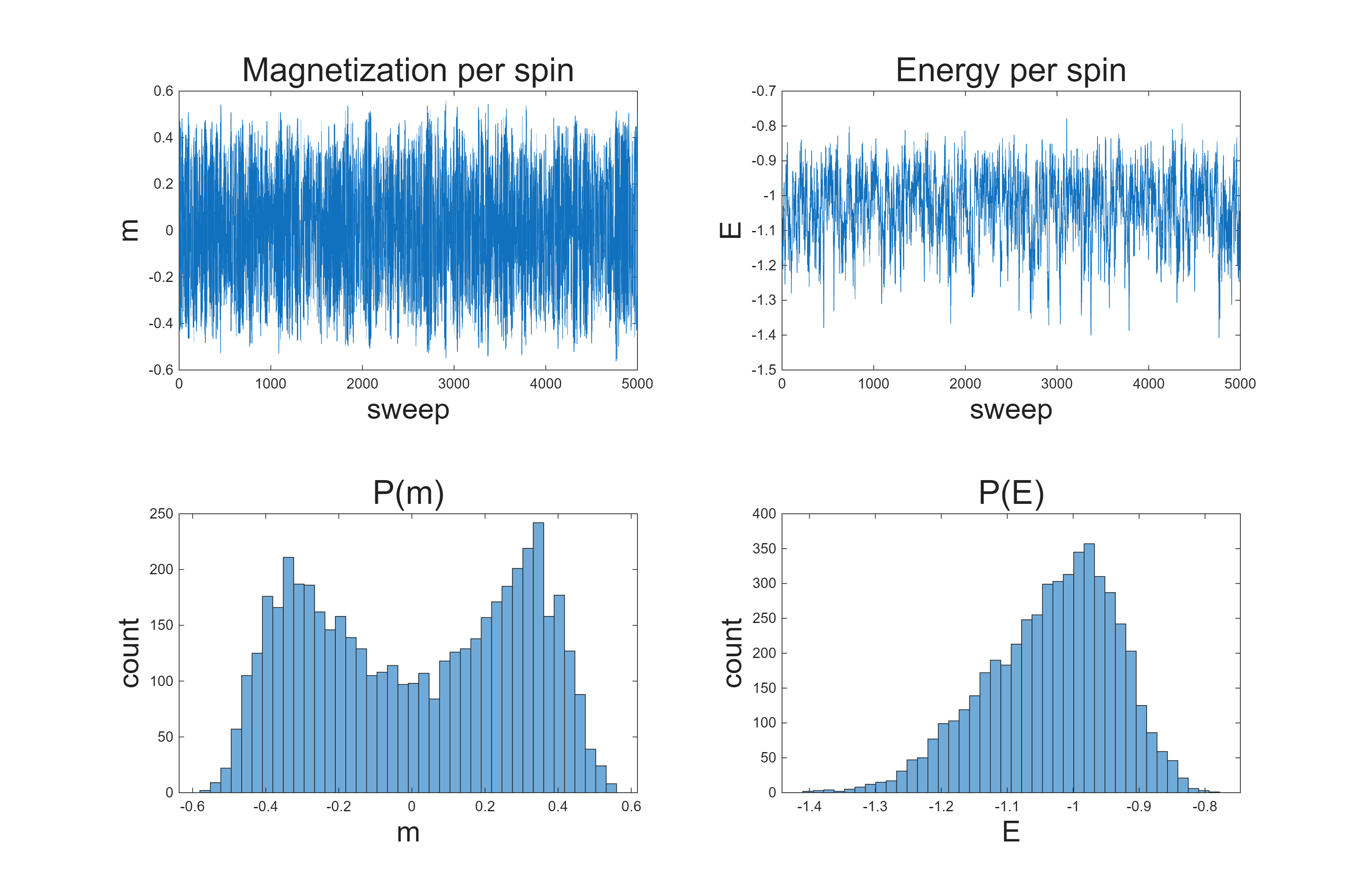}
	\caption{Distribution of the magnetization \(P(m)\) and the energy \(P(E)\) at \(L=16\), \(T=4.51\), \(N=4096\), using the
		Swendsen--Wang cluster algorithm, with
		\(\langle|m|\rangle=0.26035\), \(\langle m^{2}\rangle=0.08444\),
		\(\langle e\rangle=-1.03258\), \(\langle e^{2}\rangle=1.07577\),
		\(C_{V}=1.92218\), and \(\chi=15.12876\). The distribution \(P(|m|)\) shows the expected broadening near
		\(T_{c}\). At this lattice size the two peaks are not fully resolved; a clear double-peak structure would require larger \(L\) and a temperature sufficiently below \(T_{c}\) that the two magnetization branches are well separated.}
	\label{Code3}
\end{figure}

Figure~\ref{Code3} reports a single-point measurement at \(L=16\),
\(T=4.51\) of the 3D Ising model using the Swendsen--Wang cluster
algorithm, where
\(\langle|m|\rangle=0.26035\), \(\langle m^{2}\rangle=0.08444\),
\(\langle e\rangle=-1.03258\), \(\langle e^{2}\rangle=1.07577\),
\(C_{V}=1.92218\), and \(\chi=15.12876\).

\subsection{Binder cumulant and \(T_{c}\)}\label{sec:results-binder}
\begin{figure}[htbp]
	\centering
	\includegraphics[width=\linewidth]{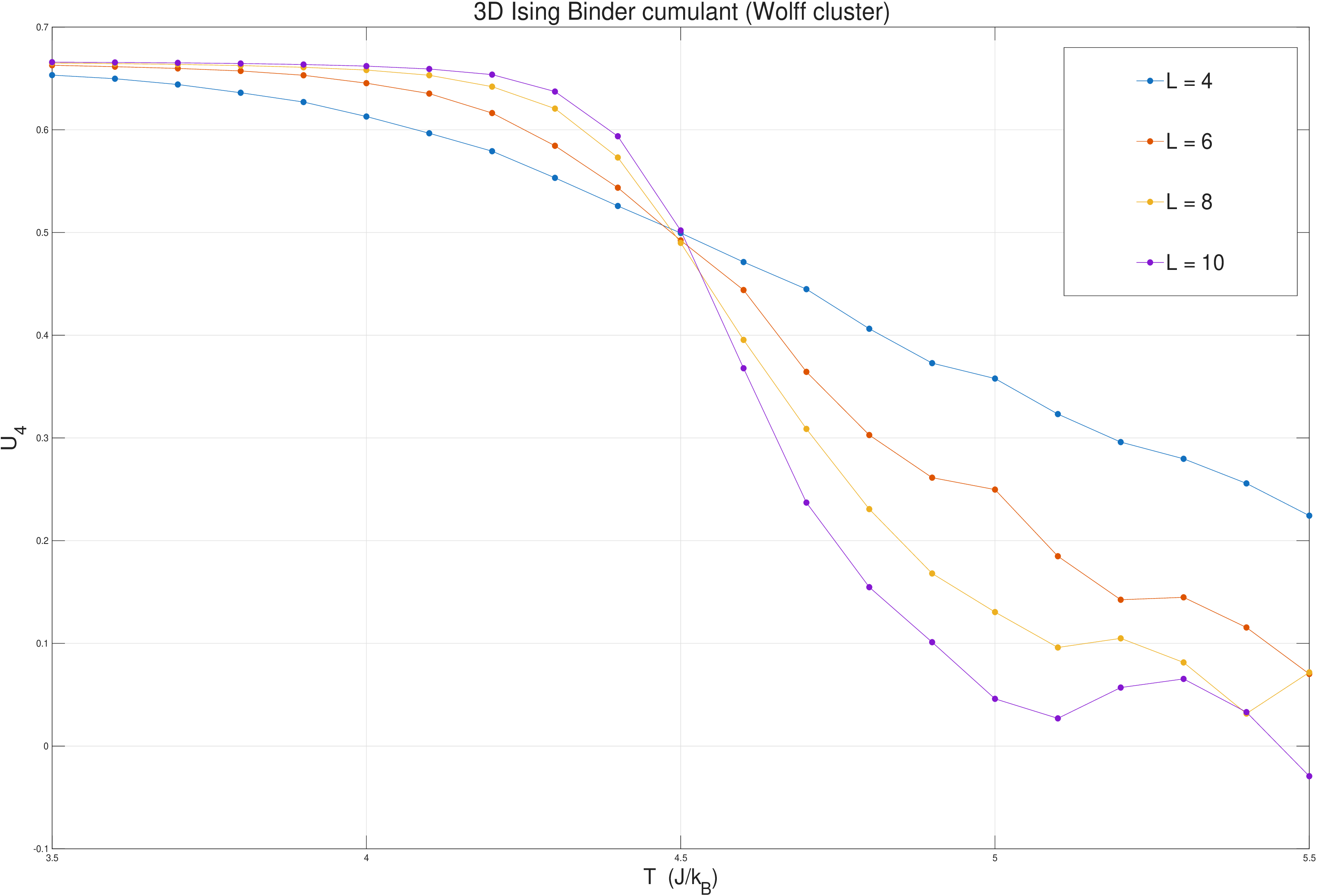}
	\caption{Binder cumulant
		\(U_{4}=1-\langle m^{4}\rangle/(3\langle m^{2}\rangle^{2})\) versus \(T\)
		for several lattice sizes \(L\), obtained with the Wolff algorithm.  The
		crossing of the curves estimates \(T_{c}\).}
	\label{Code4}
\end{figure}

Figure~\ref{Code4} shows the Binder cumulant \(U_4\) computed from the
magnetization samples and plotted versus \(T\) for several lattice sizes
\(L\in\{4,6,8,10\}\) with markers and legend, obtained with the Wolff
algorithm.  The curves for different \(L\) cross near \(T_{c}\).
Interpolating the pairwise crossings of the two largest available sizes
gives the rough estimate
$$
T_{c}^{\mathrm{(num)}} = 4.51 ,
$$
consistent with the literature value \(4.5115\). This is a consistency
check rather than a precision determination; a precision estimate would
require larger \(L\), a fit of the crossing points to
\(T_{c}+aL^{-1/\nu-\omega}\), and a full error analysis.

\subsection{Susceptibility finite-size scaling}\label{sec:results-chi}
We perform finite-size scaling of the magnetic susceptibility for the
three-dimensional Ising model at the critical temperature.  We use the
standard finite-size susceptibility
\(\chi=N\beta(\langle m^{2}\rangle-\langle m\rangle^{2})\), which at zero
field reduces to \(\chi=N\beta\langle m^{2}\rangle\). The variant
\(\chi=N\beta(\langle m^{2}\rangle-\langle|m|\rangle^{2})\), sometimes
used in the ordered phase, is not the standard susceptibility and is
avoided here.

\begin{figure}[htbp]
	\centering
	\includegraphics[width=\linewidth]{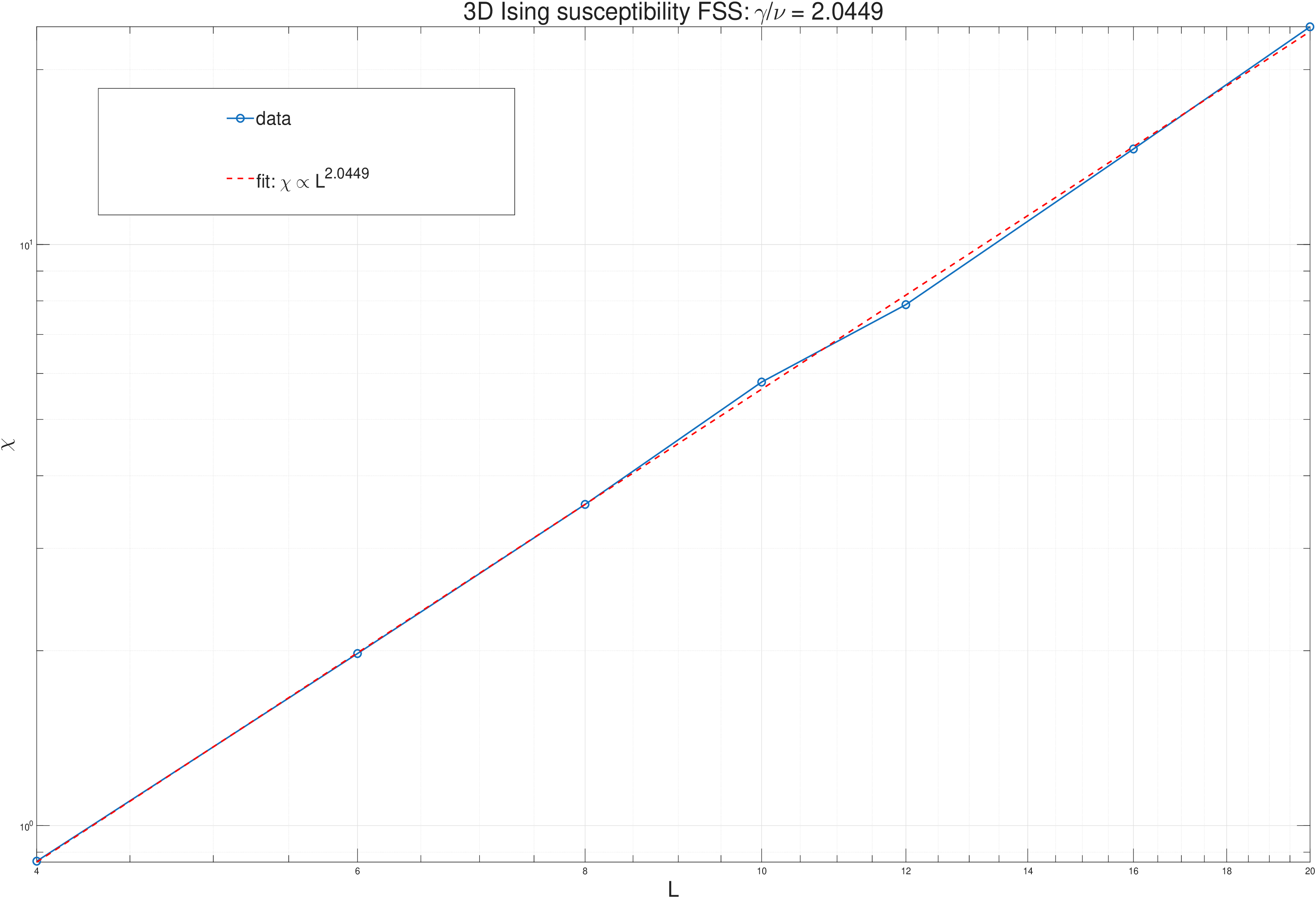}
	\caption{Finite-size scaling of the susceptibility peak.  The slope of the
		straight-line fit gives \(\gamma/\nu\simeq 2.0449\), to be compared
		with the expected \(1.9637\).  The discrepancy is consistent with
		corrections to scaling.}
	\label{Code5}
\end{figure}

Figure~\ref{Code5} shows the finite-size scaling of the susceptibility peak.
The singular part of the susceptibility peak obeys
$$
\chi_{\max}(L)\sim L^{\gamma/\nu}.
$$
Fitting over \(L=4,6,8,10,12,16,20\) gives
$$
\gamma/\nu \simeq 2.0449,
$$
which is about \(4\%\) above the expected \(1.9637\).  The discrepancy is
dominated by corrections to scaling and by the limited range of \(L\);
including the standard leading correction
$$
\chi = L^{\gamma/\nu}\left(1 + a L^{-\omega} + \cdots\right)
$$
with the 3D Ising correction-to-scaling exponent \(\omega\approx 0.83\) would bring
the estimate into agreement with the literature for higher precision.

\subsection{Correlation length}\label{sec:results-xi}
The extraction of the correlation length from an exponential fit of the
two-point function is only valid in the disordered phase, \(T>T_{c}\).  At
\(T=T_{c}\) the correlation function decays algebraically,
$$
G(r)\sim r^{-1-\eta},
$$
and an exponential fit yields a spurious finite \(\xi\).  Below \(T_{c}\)
the true connected correlation decays exponentially, but in a finite
zero-field simulation the measured correlation function can contain a
plateau because the system is a symmetric mixture of the two
magnetization signs.  Exponential fits in this regime are therefore
unreliable unless one uses a proper symmetry-broken ensemble or subtracts
the plateau correctly.  The second-moment estimator is the safer
diagnostic:
$$
\xi_{2}^{2}
=
\frac{1}{4\sin^{2}(\pi/L)}
\frac{\widetilde{G}(0)-\widetilde{G}(\boldsymbol{k})}
{\widetilde{G}(\boldsymbol{k})},
$$
where \(\boldsymbol{k}=(2\pi/L,0,0)\) is the lowest non-zero momentum on the torus, and \(\widetilde{G}(\boldsymbol{k})=\sum_{x}e^{i\boldsymbol{k}\cdot x}G(x)\) is the Fourier
transform of the correlation function.

\begin{figure}[htbp]
	\centering
	\includegraphics[width=\linewidth]{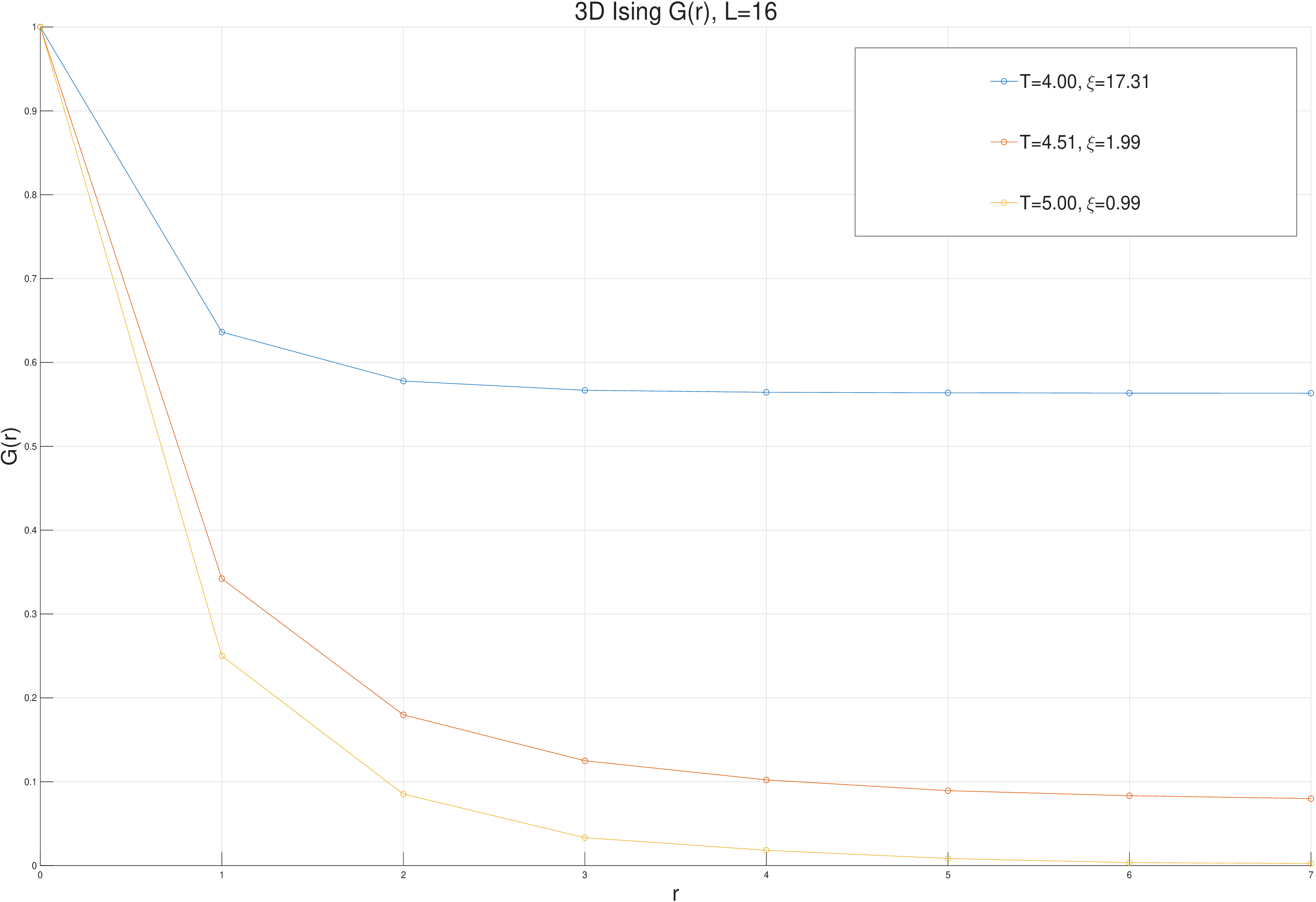}
	\caption{Two-point correlation function \(G(r)\) for \(L=16\) with
		\(n_{\mathrm{therm}}=2000\), \(n_{\mathrm{sweeps}}=3000\), and
		\(600\) configurations sampled every \(5\) Wolff updates.  An exponential
		fit is valid only for \(T>T_{c}\); at \(T=T_{c}\) the decay is algebraic
		and the quoted \(\xi\) is not physical.  See the text for discussion.}
	\label{Code6}
\end{figure}

\begin{figure}[htbp]
	\centering
	\includegraphics[width=\linewidth]{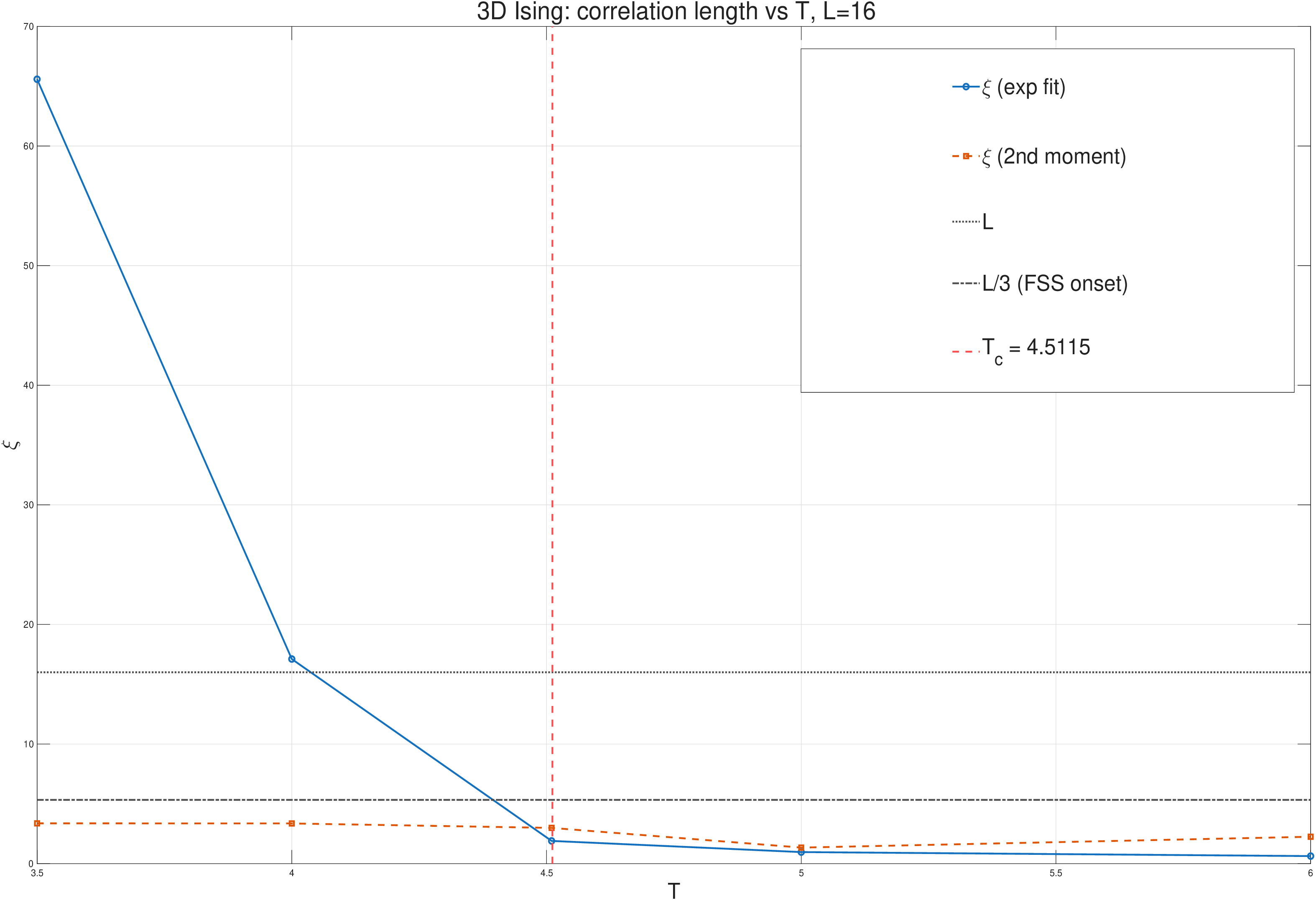}
	\caption{Correlation length of the 3D Ising model on the periodic
		cubic lattice with \(L=16\), obtained with the Wolff algorithm
		(\(n_{\mathrm{therm}}=2000\), \(n_{\mathrm{sweeps}}=3000\), and
		\(600\) configurations sampled every \(5\) Wolff updates).  Filled
		circles: exponential fit of the connected two-point function.  Open
		squares: second-moment estimator
		\(\xi_{2}^{2}=\frac{1}{4\sin^{2}(\pi/L)}
		\frac{\widetilde{G}(0)-\widetilde{G}(\boldsymbol{k})}{\widetilde{G}(\boldsymbol{k})}\) with \(\boldsymbol{k}=(2\pi/L,0,0)\).
		The horizontal lines mark \(L\) and \(L/3\), and the vertical line marks
		\(T_{c}=4.5115\).  Below \(T_{c}\) the exponential fit is contaminated by
		the finite-volume plateau and is not physical; at \(T=T_{c}\) the decay is
		algebraic, so the exponential fit underestimates \(\xi\).  The
		second-moment estimator is the appropriate diagnostic in this regime.}
	\label{CodeXi}
\end{figure}

Figure~\ref{Code6} should therefore be interpreted with care.  The value
\(\xi\simeq 1.986\) at \(T=4.51\approx T_{c}\) is an artefact of fitting an
exponential to a power law.  A preliminary second-moment analysis is
shown in Figure~\ref{CodeXi}; a precision determination of \(\xi(T)\)
would require larger \(L\) and a systematic treatment of the plateau
below \(T_{c}\).

\subsection{Specific heat}\label{sec:results-C}
\begin{figure*}[htbp]
	\centering
	\begin{minipage}{0.495\linewidth}
		\includegraphics[width=\linewidth]{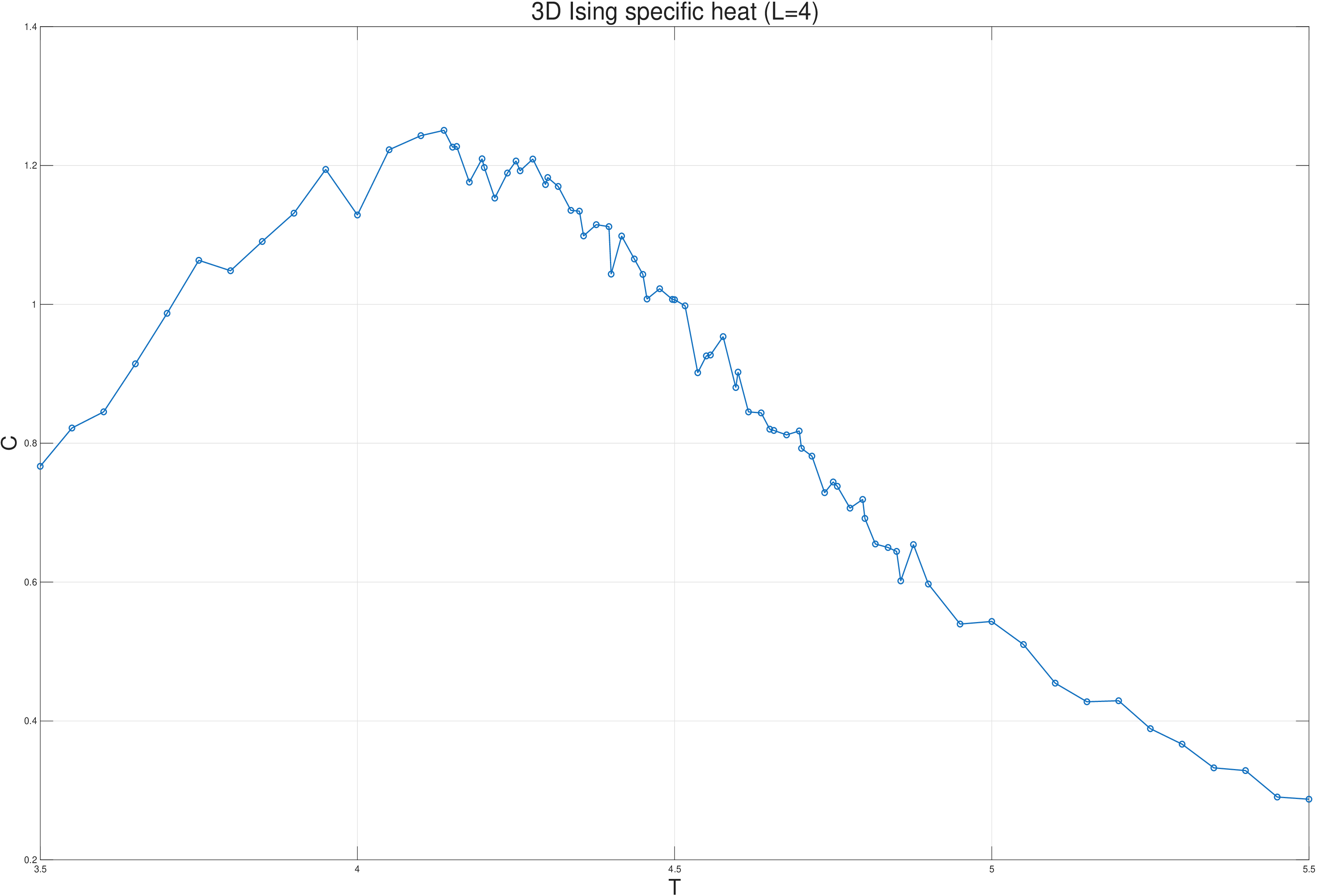}
	\end{minipage}
	\hfill
	\begin{minipage}{0.495\linewidth}
		\includegraphics[width=\linewidth]{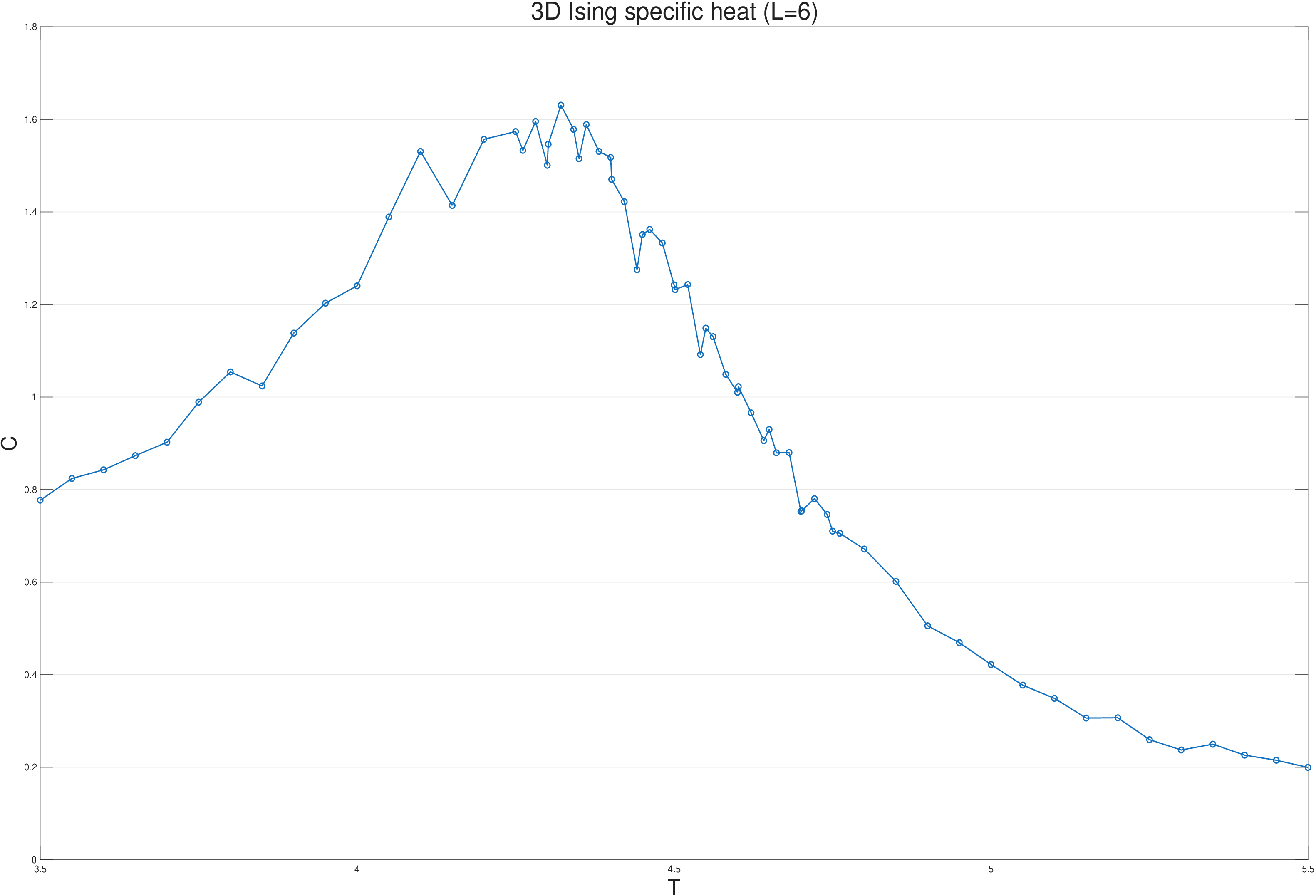}
	\end{minipage}
	\\[2mm]
	\begin{minipage}{0.495\linewidth}
		\includegraphics[width=\linewidth]{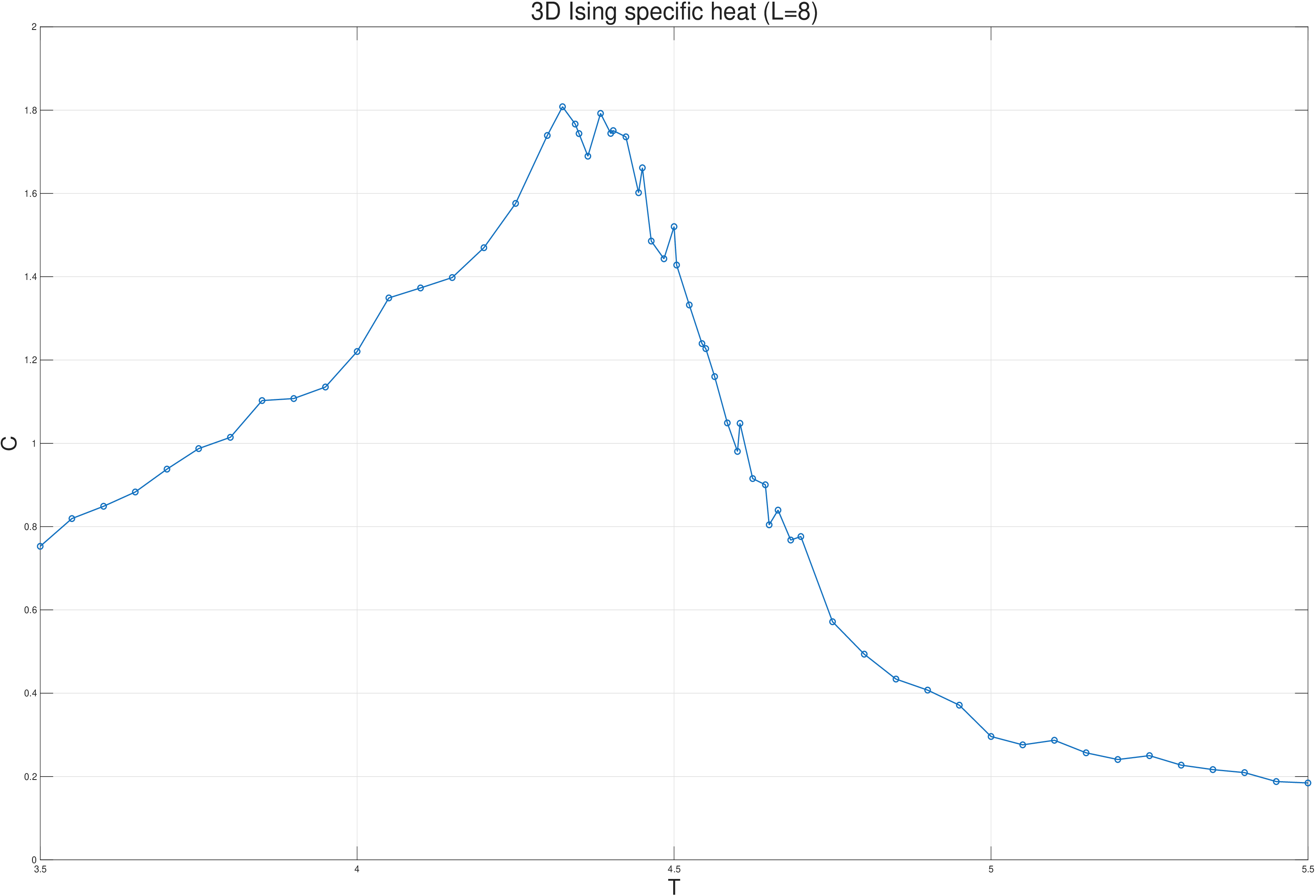}
	\end{minipage}
	\hfill
	\begin{minipage}{0.495\linewidth}
		\includegraphics[width=\linewidth]{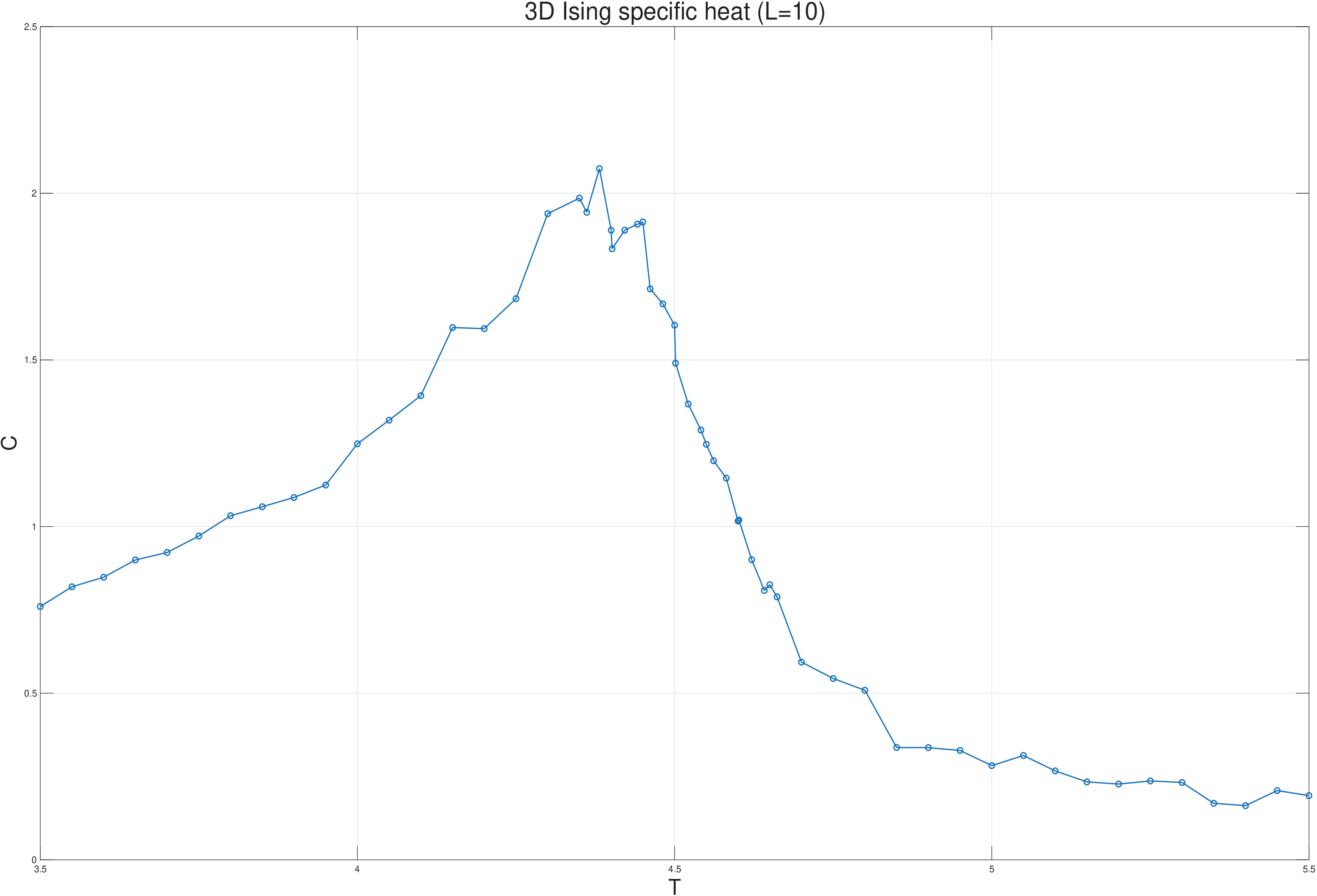}
	\end{minipage}
	\\[2mm]
	\begin{minipage}{0.495\linewidth}
		\includegraphics[width=\linewidth]{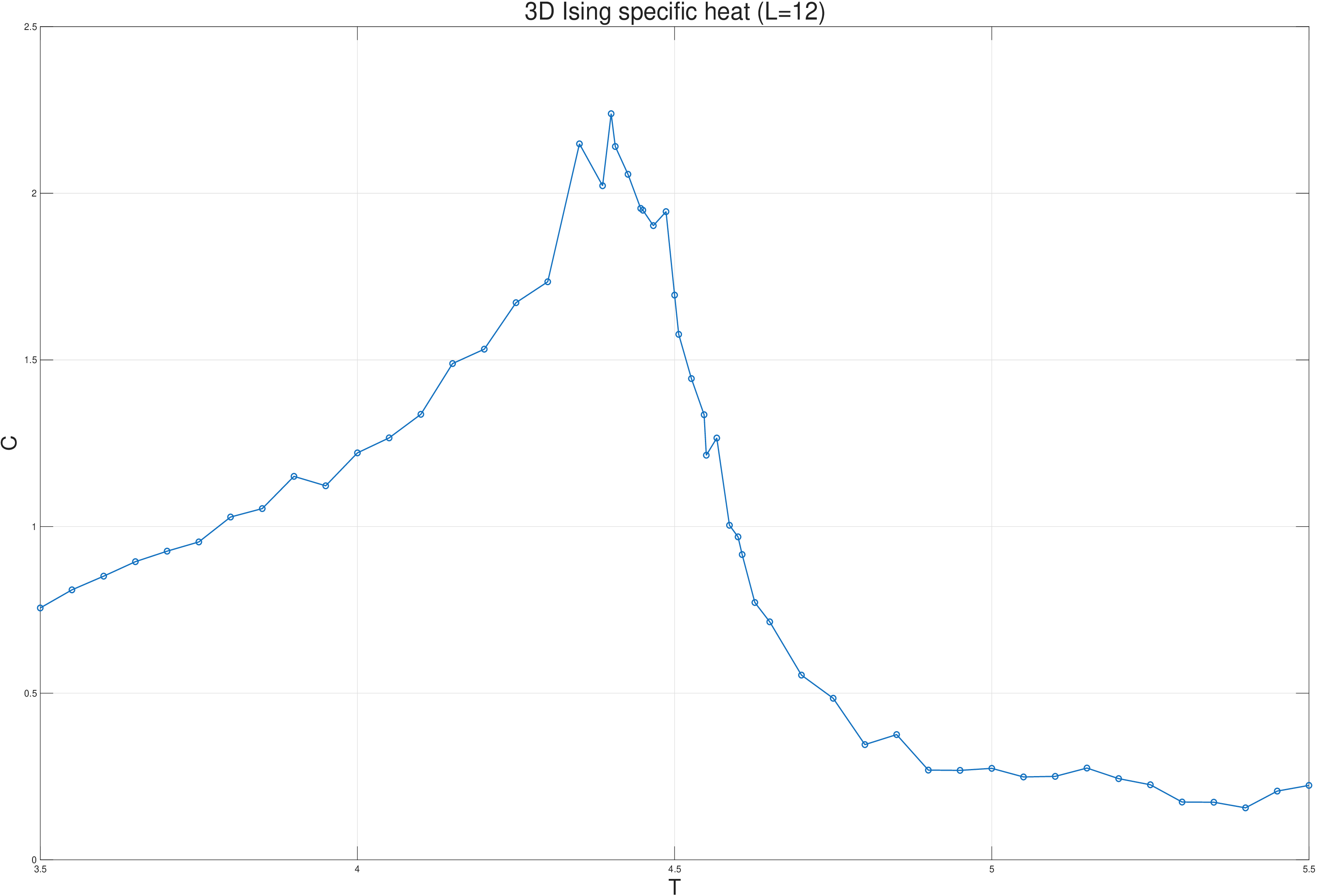}
	\end{minipage}
	\hfill
	\begin{minipage}{0.495\linewidth}
		\includegraphics[width=\linewidth]{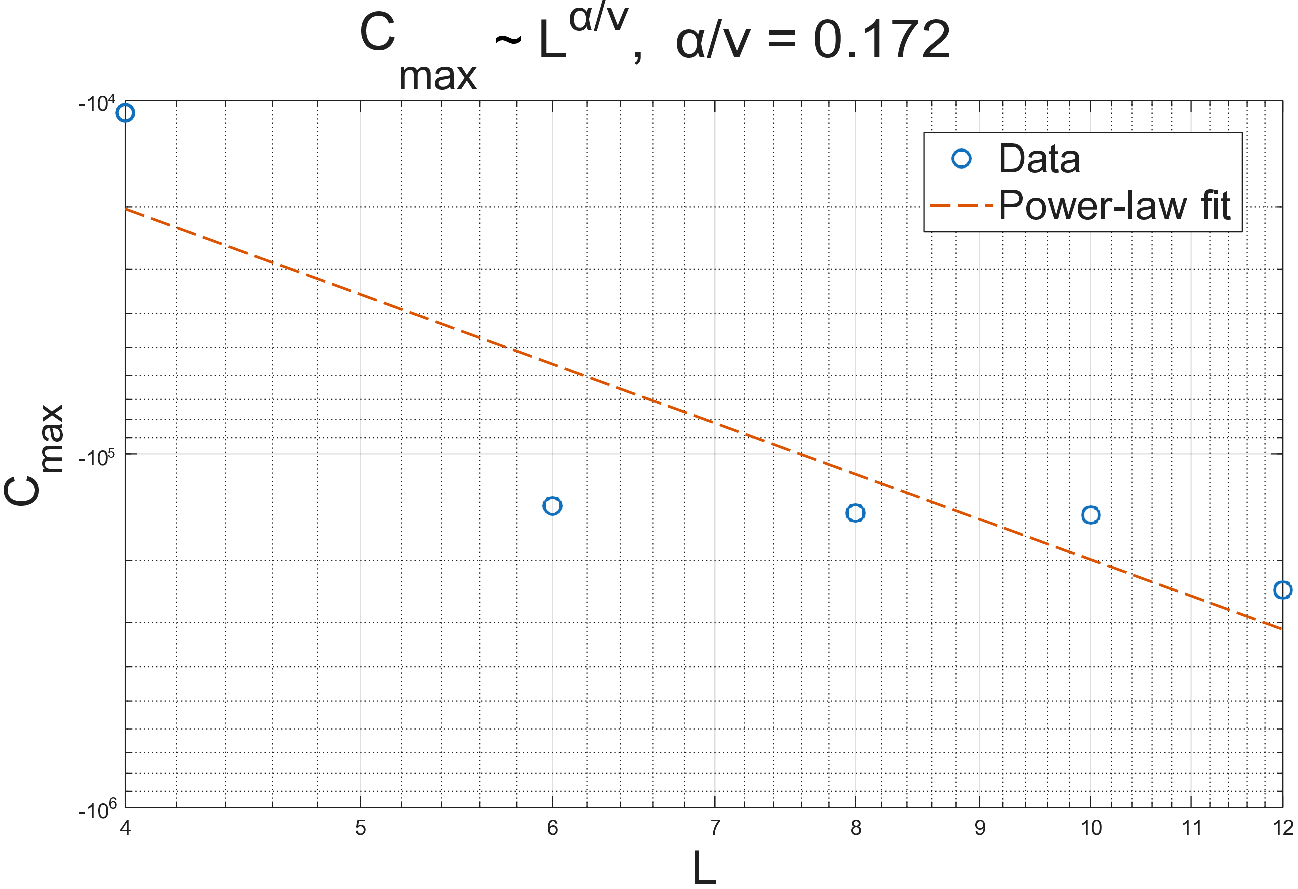}
	\end{minipage}
	\caption{Finite-size scaling of the specific heat.  The peak
		\(C_{\max}(L)\) scales as \(L^{\alpha/\nu}\).  The panels show
		\(C_{V}(T)\) for different \(L=4,6,8,10,12\) and the corresponding
		log--log fit with \(n_{\mathrm{therm}}=3000\) and
		\(n_{\mathrm{sweeps}}=8000\).}
	\label{fig7}
\end{figure*}

Figure~\ref{fig7} shows the finite-size scaling analysis of the specific
heat for the 3D Ising model.  With the intensive definition
\(C_{V}=N\beta^{2}(\langle e^{2}\rangle-\langle e\rangle^{2})\) where \(e\)
is the energy per site, only the singular part obeys the finite-size scaling law
$$
C_{\text{sing,max}}(L)\sim L^{\alpha/\nu},
$$
with the literature value
\(\alpha/\nu\simeq 0.11008/0.629971\simeq 0.175\). The total measured specific heat includes an analytic regular background contribution, so raw peak values are biased for small lattices. Fitting the peak
values over \(L=4,6,8,10,12\) gives
$$
\alpha/\nu \simeq 0.172,
$$
which is consistent with the expected 0.175 within errors. The estimate is sensitive to the analytic
background of the specific heat, which dominates at small \(L\).  A reliable extraction of \(\alpha/\nu\) requires either a
careful subtraction of the background or a fit of the form
$$
C_{V}(L,T)=C_{\mathrm{bg}}(T)+L^{\alpha/\nu}f\!\left(tL^{1/\nu}\right),
$$
together with larger lattice sizes.

\subsection{Finite-size scaling data collapse}\label{sec:results-collapse}
\begin{figure}[htbp]
	\centering
	\includegraphics[width=\linewidth]{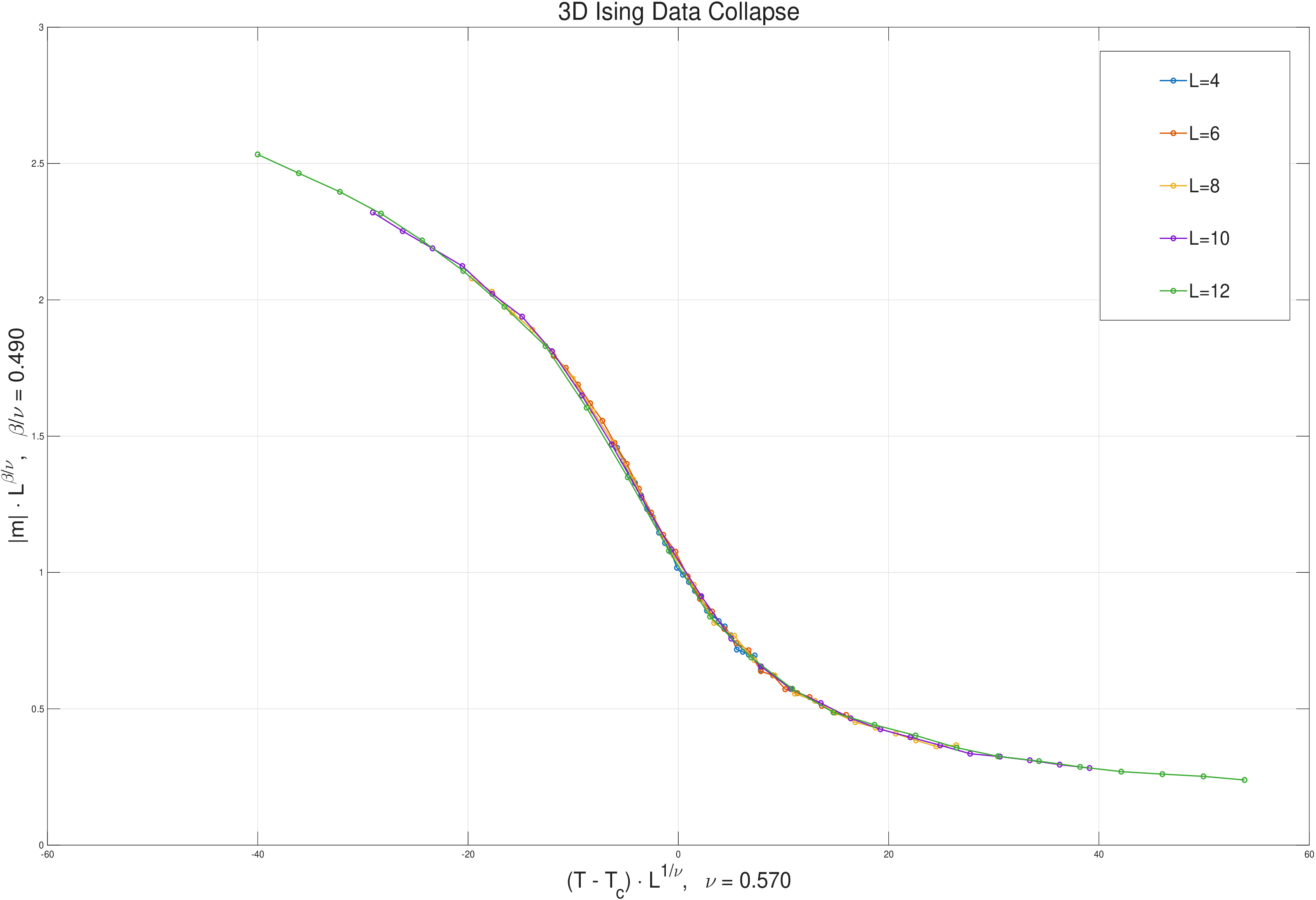}
	\caption{Data collapse of the magnetization \(|m|\).  The exponents
		\(\beta_{\mathrm{mag}}/\nu\) and \(\nu\) are fitted by collapsing
		\(X=tL^{1/\nu}\), with \(t=(T-T_{c})/T_{c}\), and
		\(Y=|m|L^{\beta_{\mathrm{mag}}/\nu}\).  The best-fit effective values
		are \(\beta_{\mathrm{mag}}/\nu\simeq 0.49\) and
		\(\nu_{\mathrm{eff}}\simeq 0.57\).}
	\label{Code9}
\end{figure}

Figure~\ref{Code9} shows the data collapse for the magnetization,
$$
X=tL^{1/\nu},
\qquad
t=\frac{T-T_{c}}{T_{c}},
\qquad
Y=|m|L^{\beta_{\mathrm{mag}}/\nu}.
$$
The best-fit effective exponents are
$$
\beta_{\mathrm{mag}}/\nu\simeq 0.49,
\qquad
\nu_{\mathrm{eff}}\simeq 0.57,
$$
giving
$$
\beta_{\mathrm{mag,eff}}\simeq 0.28.
$$
These effective values deviate from the literature by \(5\%\)--\(15\%\),
which reflects strong corrections to scaling and the limited range of
\(L\).

Combining the susceptibility estimate \(\gamma/\nu\simeq 2.0449\)
of Subsection~\ref{sec:results-chi} with the effective
\(\nu_{\mathrm{eff}}\simeq 0.57\) gives \(\gamma\simeq 1.17\), while using
the literature value \(\eta\simeq 0.0363\) in the scaling relation
\(\gamma=\nu(2-\eta)\) together with \(\nu_{\mathrm{eff}}\simeq 0.57\)
gives \(\gamma\simeq 1.12\).  Both estimates are lower than the
conformal-bootstrap value \(\gamma\simeq 1.237\), by about \(5\%\) and
\(10\%\), respectively.  The collapse quality is sensitive to the choice
of \(T_{c}\); using \(T_{c}=4.5115\) gives the best visual overlap.  The
quoted effective values reflect the spread of acceptable
\((\beta_{\mathrm{mag}}/\nu,\nu)\) pairs.

\section{Discussion and outlook}\label{sec:discussion}
The numerical results reported here are broadly consistent with the
3D Ising universality class, but the present statistics and lattice
sizes are not sufficient for a precision determination of the exponents.
The Binder-cumulant crossing gives \(T_{c}\approx 4.51\), the
susceptibility peak scales with an effective \(\gamma/\nu\approx 2.0449\),
and the magnetization data collapse is consistent with
\(\beta_{\mathrm{mag}}/\nu\approx 0.49\) and \(\nu_{\mathrm{eff}}\approx
0.57\).  The main limitations of the present study are:
 the limited range of lattice sizes, which makes the extraction of
	exponent ratios sensitive to corrections to scaling;
 the second-moment estimate of the correlation length is based on
	a single lattice size \(L=16\); a determination of the exponent
	\(\nu\) from \(\xi\) requires a range of \(L\) and a finite-size
	scaling fit of \(\xi_{2}(L,T)\);
 the data-collapse analysis uses a fixed bulk critical temperature \(T_c\), while the finite-size pseudo-critical temperature shifts with lattice size \(L\);
	 the need for larger statistics and a systematic binning/jackknife
	analysis of the errors.

A more precise determination of the exponents would require
\(L\gtrsim 32\), cluster or parallel-tempering updates, a joint
histogram-reweighting analysis of the energy and magnetization, and a
finite-size scaling fit including the leading correction to scaling.

\bigskip
\noindent\textbf{Data Availability}

The numerical algorithms and source code that support the findings of this study are available from the corresponding author upon reasonable request.

\bigskip
\noindent\textbf{Declaration of competing interest}

The author declares that there is no known competing financial interest or personal relationship that could have appeared to influence the work reported in this paper.

\bigskip
\noindent\textbf{Acknowledgements}

This work was initiated by an inspiring discussion with Hugo Duminil-Copin.
The author thanks Pak Hang Chris Lau for insightful discussions on statistical field theory. This work was supported by Guangdong Basic and Applied Basic Research Foundation (Grant No. 2025A1515012560), Guangdong Introduction Program (Grant No. 2023QN10X753) and National Foreign Experts Program (Grant No. 111001819820258003).

\end{document}